\pdfoutput=1
\documentclass[11pt,a4paper]{article} 
\usepackage[left=1.5cm,top=1cm,right=1.5cm, bottom=2cm,nohead]{geometry}
\usepackage{amsmath,amsfonts,amssymb}
\usepackage{authblk}
\usepackage{graphicx}
\usepackage{setspace}
\usepackage{tocloft}
\usepackage{soul}
\usepackage{enumerate}
\usepackage{algorithm}
\usepackage{algpseudocode}
\usepackage{subfig}
\usepackage{mathtools}
\usepackage{multirow}
\usepackage{lscape}
\usepackage{cite}
\usepackage{pbox}
\usepackage{longtable}
\usepackage{url}

\title{Cribriform pattern detection in prostate histopathological images using deep learning models}

\author[1,2,3]{Malay Singh}
\author[4]{Emarene Mationg Kalaw}
\author[5]{Wang Jie}
\author[6]{Mundher Al-Shabi}
\author[7]{Chin Fong Wong}
\author[7]{Danilo Medina Giron}
\author[8,9]{Kian-Tai Chong}
\author[6]{Maxine Tan}
\author[5]{Zeng Zeng}
\author[2,3,10,11,\footnote{Corresponding Author: Hwee Kuan Lee, leehk@bii.a-star.edu.sg}]{Hwee Kuan Lee}

\affil[1]{Computational Bioimage Analysis (CBA) Unit, Institute of Molecular and Cell Biology, Singapore}
\affil[2]{Imaging Informatics Division,Bioinformatics Institute, Singapore}
\affil[3]{ Department of Computer Science, School of Computing, National University of Singapore, Singapore}
\affil[4]{UQ Centre for Clinical Research, University of Queensland, Brisbane, Australia}
\affil[5]{Distributed Analytics Lab, Institute for Infocomm Research, Singapore}
\affil[6]{School of Engineering, Monash University Malaysia, Selangor Darul Ehsan, Malaysia}
\affil[7]{Department of Pathology, Tan Tock Seng Hospital, Singapore}
\affil[8]{PanAsia Surgery Pte Ltd, Mount Elizabeth Novena Hospital, Singapore}
\affil[9]{Surgi-TEN Specialists Pte Ltd, Farrer Park Hospital, Singapore}
\affil[10]{CNRS UMI 2955, Image \& Pervasive Access Lab ((IPAL), Singapore}
\affil[11]{Singapore Eye Research Institute, Singapore}

\date{}

\begin{document} 
\maketitle

\begin{abstract}
Architecture, size, and shape of glands are most important patterns used by pathologists for assessment of cancer malignancy in prostate histopathological tissue slides. Varying structures of glands along with cumbersome manual observations may result in subjective and inconsistent assessment. Cribriform gland with irregular border is an important feature in Gleason pattern 4. We propose using deep neural networks for cribriform pattern classification in prostate histopathological images. $163708$ Hematoxylin and Eosin (H\&E) stained images were extracted from histopathologic tissue slides of $19$ patients with prostate cancer and annotated for cribriform patterns. Our automated image classification system analyses the H\&E images to classify them as either `Cribriform' or `Non-cribriform'. Our system uses various deep learning approaches and hand-crafted image pixel intensity-based features. We present our results for cribriform pattern detection across various parameters and configuration allowed by our system. The combination of fine-tuned deep learning models outperformed the state-of-art nuclei feature based methods. Our image classification system achieved the testing accuracy of $85.93~\pm~7.54$ (cross-validated)  and $88.04~\pm~5.63$ ( additional unseen test set) across  three folds. In this paper, we present an annotated cribriform dataset along with  analysis of deep learning models and hand-crafted features for cribriform pattern detection in prostate histopathological images.
\end{abstract}

% keywords can be removed

% Include a list of up to six keywords after the abstract
%\keywords{Digital pathology, cribriform pattern detection, deep learning, prostate cancer, transfer learning}

% Include email contact information for corresponding author

%
%\begin{spacing}{2}   % use double spacing for rest of manuscript

\section{Introduction}

 The microscopic appearance of prostatic adenocarcinomas is described as having small acini arranged 
 in one or several patterns. Its diagnosis relies on a combination of tissue architectural structures 
 and cytological findings. These diagnosis criterion are considered in the 
 Gleason grading system for prostate cancer (PCa). This grading system
 is based on the glandular patterns of the tumor and is an 
 established prognostic indicator~\cite{humphrey2004gleason, gleason1977histologic}. 
 Here, various tissue architectural patterns are identified and assigned a pattern ranging from 
 1 (least aggressive) to 5 (most aggressive). Cribriform pattern in malignant glands 
 is one kind of tissue architecture in prostate, it is one of the important features 
 considered in determining if a tumor exhibits Gleason pattern 4. 
 Also, it is critical to identify Gleason 3 from Gleason 4 tumor since it changes clinical decision. 
 Only Gleason 3 lesions allow active surveillance, instead of subjecting patients to 
 surgery or radiotherapy. 

The Gleason grading system has undergone several modifications over the years~\cite{gordetsky2016grading}. 
According to several studies, cases with cribriform glands previously diagnosed as having Gleason pattern 3 would 
uniformly be considered grade 4 by today's contemporary standards~\cite{mcneal1996spread,ross2012adenocarcinomas}. 
Distinguishing whether a prostatic tumor exhibit cribriform pattern or not is relevant, since studies have reported 
that its presence in radical prostatectomy specimens are associated with biochemical 
recurrence, extraprostatic extension, positive surgical margins, distant metastases, and cancer-specific
mortality~\cite{iczkowski2011digital,kir2014association,sarbay2014significance,trudel2014prognostic,kweldam2015cribriform}.

Also, Kweldam~et al.~\cite{kweldam2015cribriform} while 
 studying the prognostic value of individual Gleason grade 4 patterns among Gleason 
 score 7 PCa patients concluded that cribriform pattern is a strong predictor 
 for distant metastasis and disease-specific death. The median time to disease-specific death in men with 
 cribriform pattern was 120 months, as compared to 150 months in men without cribriform pattern. Therefore,
 proper recognition of cribriform growth in daily pathology practice could be a useful tool
 in predicting adverse clinical outcome in PCa patients.

The Gleason grading system is inherently subjective and hence has led to high intra-observer and 
inter-observer variability. 
Various recent research contributions have suggested that 
the pathologist's training and experience affect the degree of inter-observer 
agreement~\cite{humphrey2003prostate,Allsbrook,allsbrook2001interobserver}. 
Also, diagnosis of PCa by microscopic tissue examination is tedious and time 
consuming. 

The aforementioned issues of low inter-observer agreement and the requirement of identifying 
various types of glandular patterns has motivated research for development of automated image 
based grading systems for PCa. Various computer-aided diagnosis (CAD)
systems have been developed using a multitude of machine learning, image processing,
and feature extraction methods~\cite{madabhushi2016image, nir2019comparison}.
These systems have usually automated the task(s) of 
object detection, image/object classification, and 
image segmentation for aiding pathologists. For PCa, CAD systems have generally emphasized on
gland segmentation, nuclei segmentation, and image classification tasks. Cribriform pattern classification
 is a different task for the conventional PCa CAD systems and it is yet to get the much needed attention. 
 This paper is an attempt to fill in this gap by 
 presenting an automated image based cribriform pattern classification system.  
 The main contribution of this paper are  
 \begin{enumerate}
 \item 
 our annotated cribriform dataset,
 \item  hand-crafted nuclei features,  and  
 \item combination of nuclei features with deep learning (DL) models

 \end{enumerate}
 for cribriform pattern detection 
 in prostate histopathological images.

 These hand-crafted nuclei 
 features are designed to incorporate relevant nuclei 
 texture and spatial information for cribriform pattern detection. The DL architectures used in our method have been chosen 
 and/or modified according to their performance in similar histopathological 
 tasks as suggested in literature~\cite{shin2016deep,coudray2018classification,sharma2017deep,bejnordi2018using,gecer2018detection,araujo2017classification,alom2019advanced}. 
Recently, various deep models like ResNet~\cite{he2016ResNet}, VGG16~\cite{simonyan2014very}, VGG19~\cite{simonyan2014very}, 
 Inception-v3 (GoogLeNet)~\cite{szegedy2015going,szegedy2016rethinking},  and DenseNet~\cite{huang2017densely}
  have achieved top performance in the ImageNet~\cite{russakovsky2015imagenet} challenge. 
% Medical images with their heterogeneous patterns have warranted the need of a more sophisticated DL model when compared to natural images. 
 This paper builds upon the recent success of DL in medical images' 
 tasks~\cite{shin2016deep,coudray2018classification,sharma2017deep,bejnordi2018using,gecer2018detection,araujo2017classification} and 
 robust performance of ResNet~\cite{he2016ResNet}, VGG16~\cite{simonyan2014very}, VGG19~\cite{simonyan2014very}, 
 Inception-v3 (GoogLeNet)~\cite{szegedy2015going,szegedy2016rethinking},  and DenseNet~\cite{huang2017densely} for the task of cribriform pattern detection. 
 These DL architectures have been fine-tuned via transfer learning before combination with hand-crafted nuclei 
 features for cribriform pattern detection. This paper focuses on the clinical problem of cribriform pattern detection and provides promising machine learning based method to aid pathologists.

\section{Related work}
\label{sec:related_work}

Various CAD systems have been developed for 
prostate histopathological image classification while automating 
gland segmentation, nuclei segmentation, and image classification 
tasks~\cite{madabhushi2016image,xu2010high,nguyen2014prostate,kwak2017multiview,doyle2007automated, kwak2017nuclear, niazi2017visually, Diamond,khan2017predicting,gummeson2017automatic,kallen2016towards,litjens2017survey,yap2015automated,singh2017gland,singh2017study,ali2013cell,nir2019comparison}.
Cribriform pattern classification is an
altogether new task for the conventional PCa CAD systems. 
%We discuss a few 
%PCa CAD systems for possible ideas w.r.t. automated cribriform pattern classification system. 
A general pipeline for prostate histopathological image classification 
is gland segmentation followed by feature extraction from these segmented glands for 
classification~\cite{xu2010high,nguyen2014prostate,kwak2017multiview}. 

Few approaches like Diamond~et al.~\cite{Diamond} and 
Lin~et al.~\cite{lin2016curvelet} have bypassed this segmentation step.
Diamond~et al.~\cite{Diamond} proposed using morphological 
and textural features to identify regions belonging to 
 stroma, PCa, and normal tissue. 
 Lin~et al.~\cite{lin2016curvelet} used curvelet-based textural features with 
 Support Vector Machine (SVM)~\cite{svm} for classifying 
a given prostate histopathological image as Gleason patterns 3+3, 3+4, 4+3, and 4+4.

Nguyen~et al.~\cite{nguyen2014prostate} used shape and textural features to identify nuclei regions. 
A nuclei-lumen graph made from nuclei and lumen boundary pixels
 was processed by normalized cuts~\cite{shi2000normalized} for final gland
segmentation. This paper then used various 
graph based features with SVM~\cite{svm} for automated PCa grading.
Kwak~et al.~\cite{kwak2017multiview} proposed using multiple scales
 in the same system for PCa grading. 
% Kwak~et al.~\cite{kwak2017nuclear} proposed using for PCa grading. 
Nuclei, gland and lumen regions were segmented using features in HSV and CIELab color spaces. For a given image, first the 
segmentation was performed and morphological features at multiple scales
were used for final automated PCa grading. 
In an another similar approach, Ali~et al.~\cite{ali2013cell} proposed using nuclei-graphs to compute features for
predicting biochemical recurrence in prostate histopathological tissue microarray images. 
Fukuma~et al.~\cite{fukuma2016study} and Khan~et al.~\cite{khan2017predicting} also proposed using 
nuclei graph features for automated grading of brain and prostate histopathological images respectively.

The methods as discussed above focused on the development of hand-crafted features
which are to be used along with classical machine learning methods. 
They also focused on a different problem of prostate histopathological 
image classification instead of cribriform pattern classification. 
On similar lines, various DL architectures have been deployed for 
prostate histopathological 
images' tasks\cite{greenspan2016guest,madabhushi2016image,litjens2016deep,kwak2017nuclear,alom2019advanced}.
Generally, DL architectures require a preferably large dataset for training and evaluation purposes
due to their huge parameter space.
As manually annotated data in the 
medical imaging domain is scarce, various recent research efforts 
have focused on transfer learning~\cite{shin2016deep,chang2017unsupervised,gessert2019deep,swati2019brain,khan2019novel,hekler2019pathologist,brancati2019deep,ahmad2019classification,hosny2019classification,rai2019investigation,kather2019deep}. 
One of the approach for transfer learning is fine-tuning of pre-trained DL networks.  
In fine-tuning of pre-trained network, some layers  are frozen during training along with small learning rate. 
We list out a few recent approaches with the corresponding pre-trained models 
used via fine-tuning along with medical image task as follows:

\begin{itemize}
\item Shin~et al.~\cite{shin2016deep}: Uses GoogLeNet~\cite{szegedy2015going} and AlexNet~\cite{krizhevsky2009learning} 
for ``Thoracoabdominal Lymph Node Detection'' and ``Interstitial Lung Disease Classification''.

\item Gessert~et al.~\cite{gessert2019deep}: Uses  ResNet~\cite{he2016ResNet}, VGG16~\cite{simonyan2014very}, and DenseNet~\cite{huang2017densely} for cancer tissue identification in confocal
laser microscopy images for colorectal cancer.

\item Khan~et al.~\cite{swati2019brain}: Uses VGG16~\cite{simonyan2014very} for brain tumor classification in 
Magnetic Resonance (MR) images. 

\item Khan~et al.~\cite{khan2019novel}: Uses GoogLeNet~\cite{szegedy2015going}, ResNet~\cite{he2016ResNet}, and VGG16~\cite{simonyan2014very} for breast cancer cytological image classification. They also combined these fine-tuned networks by average pooling. 

\item Hekler~et al.~\cite{hekler2019pathologist}: Uses ResNet~\cite{he2016ResNet} for H\&E  stained melanoma histopathological image classification. 

\item Brancati~et al.~\cite{brancati2019deep}: Uses ResNet~\cite{he2016ResNet}  for  invasive ductal carcinoma  detection and lymphoma classification. 
 
\item Ahmad~et al.~\cite{ahmad2019classification}: Uses ResNet~\cite{he2016ResNet}, GoogLeNet~\cite{szegedy2015going}, and AlexNet~\cite{krizhevsky2009learning}  for breast cancer cytological image classification.
 
\item Hosny~et al.~\cite{hosny2019classification}: Uses AlexNet~\cite{krizhevsky2009learning}  for skin lesion  image classification.

\item Kather~et al.~\cite{kather2019deep}: Uses ResNet~\cite{he2016ResNet} to predict microsatellite instability in gastrointestinal cancer.
 
 \end{itemize}

Apart from the latest transfer learning based CAD approaches, various DL
architectures have also been used for breast cancer and lung cancer histopathological images. 
Coudray~et al.\cite{coudray2018classification} trained an 
Inception-v3 (GoogLeNet)~\cite{szegedy2015going,szegedy2016rethinking} on whole slide images (WSI) obtained from The Cancer Genome 
Atlas to automatically classify histopathology images into Adenocarcinoma (LUAD), squamous cell carcinoma (LUSC) or normal lung tissue. 
Sharma et al.\cite{sharma2017deep} studied H\&E stained histopathological images of gastric carcinoma and applied deep learning to 
 classify cancer based on immunohistochemical response and necrosis detection based on the existence of tumor necrosis in the tissue. 
Bejnordi et al.\cite{bejnordi2018using} applied deep learning on $2387$ H\&E stained breast cancer 
images to discriminate between stroma surrounding 
invasive cancer and stroma from benign biopsies. Gecer~et al.~\cite{gecer2018detection} proposed an algorithm based on deep 
convolutional networks that classify 
WSI of breast biopsies into five diagnostic categories. Araujo~et al.~\cite{araujo2017classification} designed a multi-scale 
deep convolutional neural network to 
classify normal tissue, benign lesion, in situ carcinoma, and invasive carcinoma, and in two classes, carcinoma and non-carcinoma.

Various recent approaches in machine learning literature
have suggested using deeper networks for better classification/detection performance
~\cite{he2016ResNet,huang2017densely}. 
 Following which, various deep models like ResNet~\cite{he2016ResNet} and DenseNet~\cite{huang2017densely}
have achieved top performance in the 
 ImageNet~\cite{russakovsky2015imagenet} challenge. These networks 
 have outperformed the previous top performer GoogLeNet~\cite{szegedy2015going}.
On the other hand, medical images with their heterogeneous patterns has warranted need of a 
more sophisticated DL model when compared to natural images. This paper builds upon the 
 recent success of DL in 
 medical images' tasks and top performance of 
ResNet~\cite{he2016ResNet} and DenseNet~\cite{huang2017densely} for the task of cribriform pattern classification. 
 These two networks have been compared with SVM classifier which used 
 nuclei based features~\cite{fukuma2016study, kwak2017nuclear,khan2017predicting},
 VGG16~\cite{simonyan2014very}, VGG19~\cite{simonyan2014very}, and 
 Inception-v3 (GoogLeNet)~\cite{szegedy2015going,szegedy2016rethinking}. 
 The VGG16~\cite{simonyan2014very}, VGG19~\cite{simonyan2014very}, and 
 Inception-v3 (GoogLeNet)~\cite{szegedy2015going,szegedy2016rethinking} are some of the initial
 DL architectures which achieved high performance across large scale 
 natural image datasets.
 In this paper, the performance of 
`ResNet-50' which is ResNet~\cite{he2016ResNet} network with 50 layers 
along with `DenseNet-121', `DenseNet-169' which are DenseNet~\cite{huang2017densely} 
networks with 121 and 169 layers respectively are studied for the task of 
cribriform pattern detection. All these DL architectures have been fine-tuned via transfer learning. 
The fine-tuned DL architectures are then combined with hand-crafted nuclei features using Multi-layer 
Perceptron (MLP) for our final results. This paper focuses on the clinical problem of cribriform pattern detection and 
provides promising machine learning based method to aid pathologists. 

\section{Dataset}
\subsection{Dataset preparation}

H\&E stained whole slide images were downloaded from the `Legacy Archives' of the 
NCI Genomic Data Commons (GDC)~\cite{gdc}. 
The GDC Legacy Archives currently hosts much of 
``The Cancer Genome Atlas (TCGA)''~\cite{tc} data.
 The TCGA has various WSIs categorised according to cancer types. 
 Each WSI has a unique patient ID (Slide Name) 
 in TCGA. This patient information is important when we design the experiments. The design 
 should be such that patients sets among training, testing, and validation sets are mutually 
 exclusive for reliable experiments and results.

\begin{figure}[htbp]
\begin{center}
\fbox{\includegraphics[width=\textwidth]{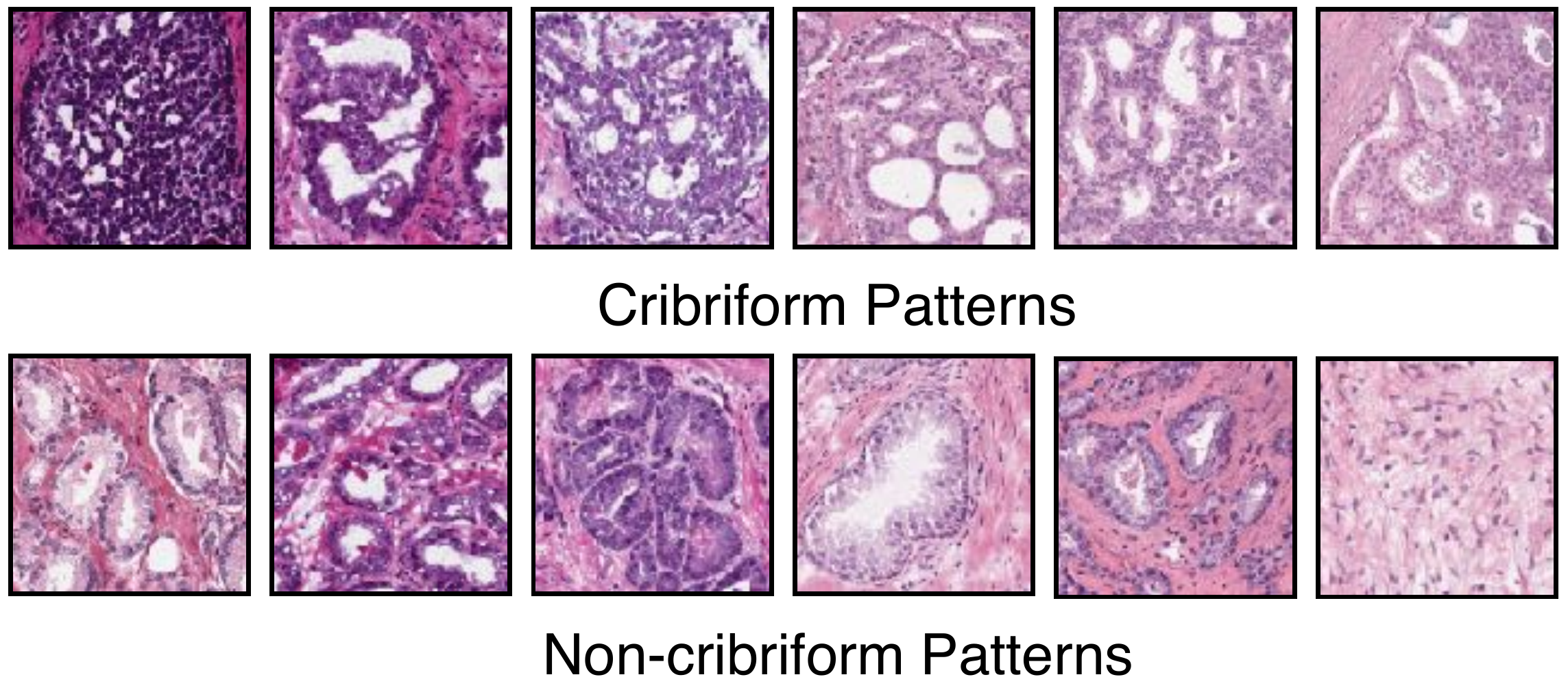}}
\caption{Example H\&E images with `Cribriform' and `Non-cribriform' patterns in our dataset. 
These images were extracted at $40\times$ magnification with pixel resolution of 0.25MPP.
The cribriform pattern detection system was developed using H\&E images with different color variations.}
\label{fig:sample_images_in_dataset}
\end{center}
\end{figure}

Cribriform pattern may be seen in both benign 
and malignant glands. Neoplastic cribriform gland pattern may be seen in high grade prostate intraepithelial neoplasia (HG-PIN),
 acinar adenocarcinoma Gleason pattern 4, intraductal carcinoma of the prostate (IDC-P), and prostatic duct adenocarcinoma. 
Some example images for both `Cribriform' and `Non-cribriform' patterns are illustrated in Fig.~\ref{fig:sample_images_in_dataset}.
 Cribriform patterns are characterized by solid proliferation with multiple 
 punched-out lumina, without intervening stroma~\cite{kweldam2015cribriform} 
 as evident in the first row of Fig.~\ref{fig:sample_images_in_dataset}.

 Usual approach of data preparation is a pathologist 
 going through the WSI using Aperio ImageScope~\cite{imagescope} 
 and then extract images containing regions of interest(ROIs). These ROIs
 will either contain a cribriform pattern or a Non-cribriform pattern 
 and hence labelled accordingly. 
 We followed this protocol and initially 
 extracted 161 images ($1024 \times 1024$ pixels) at $40\times$ from WSI of 10 patients  using 
 Aperio ImageScope~\cite{imagescope}.
The $1024 \times 1024 $ pixels dimension was chosen by the pathologist 
 such that the corresponding field of view contained enough information to identify if the image contains a
 cribriform pattern or not.
 The subsequent experiments for cribriform detection using deep learning were inconclusive due to 
 insufficient patient data. We 
 then extracted $3072 \times 3072$ pixels images from 9 more patients using 
 Aperio ImageScope~\cite{imagescope} and OpenSlide~\cite{goode2013openslide}. 
 These images were then annotated by 
the pathologists in our team as `Cribriform' or `Non-cribriform'. Table~\ref{tab:cribriform_dataset} tabulates the 
 number of manually extracted and annotated images from each patient.
 This way we extracted 728 labeled images from 19 patients. 
 Apart from these labeled images there were some images which 
 were rejected during the labelling process as they were ambiguous and/or tissue structure was not preserved well. 
% The images were labeled as `Reject' and excluded from current experiments. 
 
 After manually going through the images with the pathologist for labelling individual images
 we augmented the data using 
translation and rotation operations. The following section describes the data augmentation process.

\begin{table}[htbp]
\centering
	\caption{Description of the \textbf{manually extracted and annotated} 
	images in the cribriform dataset. 
	We have 12 unique cribriform and 7 unique 
	non-cribriform patients. %Ideally, we need 9 unique cribriform and 9 unique non-cribriform patients for our experiments.
	}
	\label{tab:cribriform_dataset}

\begin{tabular}{|p{0.8cm}|p{3.2cm}|p{3.5cm}|p{2.3cm}|p{1.8cm}|p{3.3cm}|} \hline
\textbf{S.N.}&\textbf{Slide Name \quad (Patient ID)} &\textbf{Gleason grade} &\textbf{Number of Cribriform Images} &\textbf{Number of Non-cribriform Images} &\textbf{Image Dimensions}\tabularnewline \hline
1 & TCGA-2A-A8VO & 3+3 (HG-PIN) & - & 17 & $1024 \times 1024$ pixels\\ \hline
2 & TCGA-2A-A8VT & 3+3 (HG-PIN) & 2 & -& $1024 \times 1024$ pixels\\ \hline
3 & TCGA-EJ-5510 & 4+3 (HG-PIN) & 6 & 1& $1024 \times 1024$ pixels\\ \hline
4 & TCGA-EJ-5511 & 3+4 (HG-PIN) & 1 & 16& $1024 \times 1024$ pixels\\ \hline
5 & TCGA-EJ-5519 & 4+4 (HG-PIN) & 5 & -& $1024 \times 1024$ pixels\\ \hline
6 & TCGA-EJ-7797 & 3+4 (HG-PIN) & - & 21& $1024 \times 1024$ pixels\\ \hline
7 & TCGA-G9-6338 & 4+3 (No HG-PIN) & - & 36& $1024 \times 1024$ pixels\\ \hline
8 & TCGA-G9-6363 & 4+3 (HG-PIN) & - & 14& $1024 \times 1024$ pixels\\ \hline
9 & TCGA-HC-7211 & 3+4 (HG-PIN) & 25 & -& $1024 \times 1024$ pixels\\ \hline
10 & TCGA-HC-7212 & 3+4 (HG-PIN) & 17 & -& $1024 \times 1024$ pixels\\ \hline \hline
11 & TCGA-EJ-7791 & No report & 1& 51 & $3072 \times 3072$ pixels\\ \hline
12 & TCGA-EJ-8469 & 4+5 (HG-PIN) & 121 & - & $3072 \times 3072$ pixels\\ \hline
13 & TCGA-EJ-A46F & 4+4 (HG-PIN) & 86 & - & $3072 \times 3072$ pixels\\ \hline
14 & TCGA-FC-7708 & No report & 5 & 60 & $3072 \times 3072$ pixels\\ \hline
15 & TCGA-HC-7078 & No report & 1 & 12 & $3072 \times 3072$ pixels\\ \hline
16 & TCGA-HC-7820 & 3+4 (HG-PIN) & - & 9 & $3072 \times 3072$ pixels\\ \hline
17 & TCGA-XJ-A9DI & 5+4 (No HG-PIN) & - & 28 & $3072 \times 3072$ pixels\\ \hline
18 & TCGA-XK-AAJP & 4+3 (HG-PIN) & -& 80 & $3072 \times 3072$ pixels\\ \hline
19 & TCGA-YL-A8HL & 4+5 (No HG-PIN) & 114 & - & $3072 \times 3072$ pixels\\ \hline
%20 & TCGA-HC-A631 & 4+5 (HG-PIN) & - & - & $3072 \times 3072$ pixels\\ \hline
%21 & TCGA-KK-A6E1 & 4+5 (No HG-PIN) & -&- & $3072 \times 3072$ pixels\\ \hline
%22 & TCGA-KK-A8IA & No report (To grade) & - & - & $3072 \times 3072$ pixels\\ \hline
%23 & TCGA-XJ-A9DK & 4+4 (No HG-PIN) & - & - & $3072 \times 3072$ pixels\\ \hline
%..& \multicolumn{5}{|c|}{\textbf{Malay will extract more patient slides if needed from TCGA/GDC.}} \\ \hline
\multicolumn{3}{|c|}{\textbf{Total $1024\times 1024$ pixels images from 10 patients}} & \textbf{56} & \textbf{105} & \\ \hline
\multicolumn{3}{|c|}{\textbf{Total $3072\times 3072$ pixels images from 9 patients}} & \textbf{328} & \textbf{240} & \\ \hline
\multicolumn{3}{|c|}{\textbf{Total (749 images from 19 patients)}} & \textbf{384} & \textbf{365} & \\ \hline
\end{tabular}
\end{table}

 \subsection{Data augmentation}
 \label{sec:data_augmentation}
 
\begin{figure}[htbp]
\begin{center}
\fbox{\includegraphics[width=0.5\textwidth]{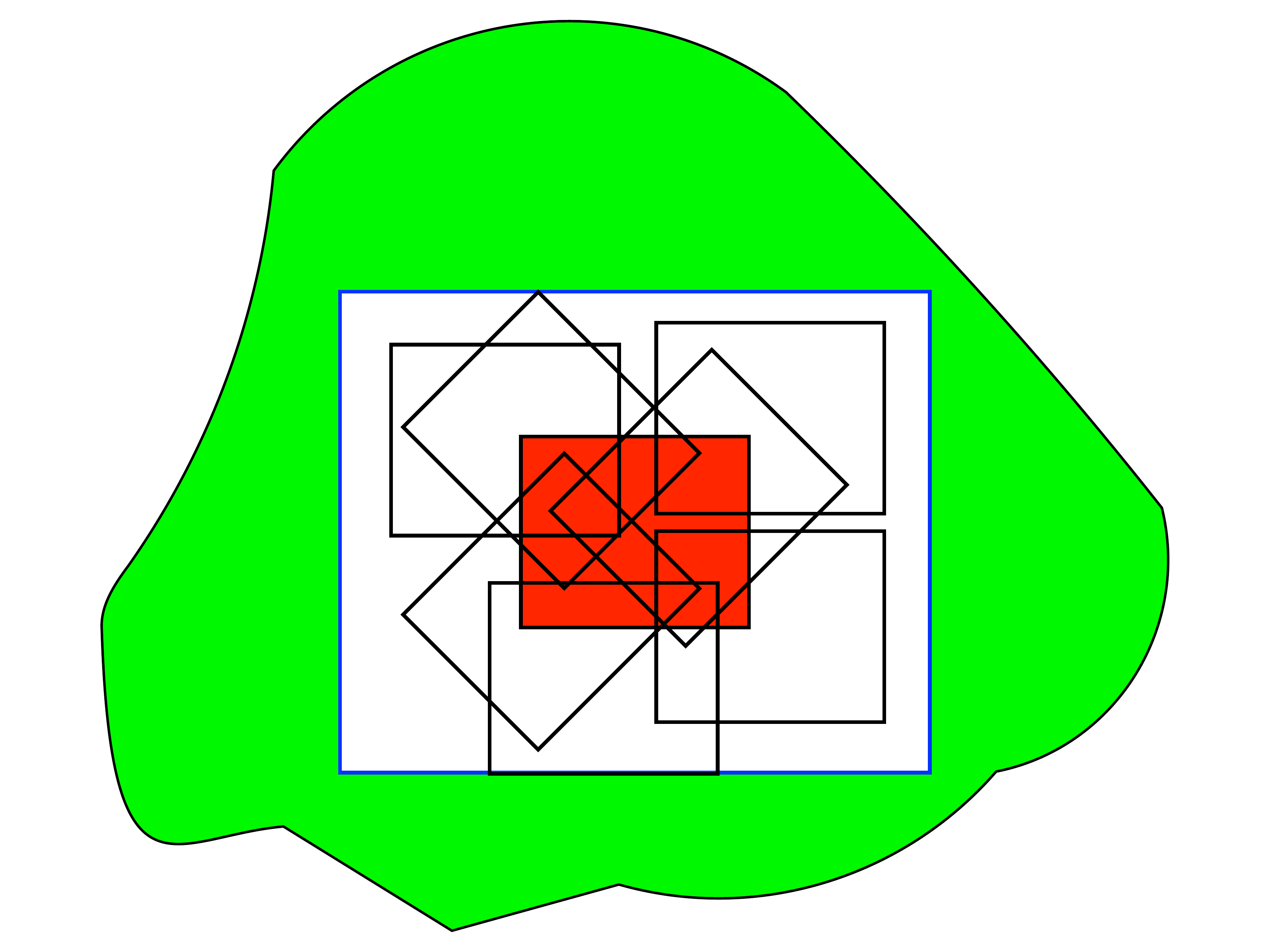}}
\caption{Example of extraction of new images from Whole Slide Image (WSI). A WSI is indicated as an arbitrary 
structure filled with green. The originally extracted $1024\times 1024$ pixels image is indicated by a filled red box. The surrounding 
$5000 \times 5000$ pixels region is indicated by blue bordered box filled with white. The new images are to be sampled from inside 
this region. Some of the sampled images after rotation and/or translation from the original image are indicated by black empty boxes.
Translation can be done by $50$ and $100$ pixels in horizontal and vertical directions from a given image. 
From original location of $(x_c,y_c)$ to 
$(x_c\pm k\Delta, y_c\pm k\Delta)$ locations where $k\in \{0,1,2\}$ and $\Delta=50$. We can extract more images after 
 rotations of $60^\circ$ and $120^\circ$ from a given image location. This will give us 
 $75(=5\times 5\times 3)$ times the original dataset of images.}
\label{fig:sampling}
\end{center}
\end{figure}

 We augment the dataset by using translation, rotation based sampling in the WSI. 
 Given that we know the location of extracted $1024\times 1024$ and $3072 \times 3072$ pixels 
 images in the WSI, we can extract a region of around $5000\times 5000$ pixels around it using OpenSlide~\cite{goode2013openslide}. 
 In this extracted region we can sample new images by translation of 50-100 pixels to 
 the left, right, top, bottom of from the position of original image. Apart from translation, we can also sample images by 
 rotation with and without translation. Fig.~\ref{fig:sampling} illustrates the idea this idea for data augmentation. 
 The images which are extracted around a given unique location will have same label as the original image location. 
 %Once we have sampled the images in above fashion, we can then ask Ema to confirm the annotation. 
 
 Let us define the total number of rotation, translations for extraction of new images. 
 This will aid us in estimating the size of the augmented 
 dataset. We define the translations of $50(=\Delta)$ and $100(=2\Delta)$ pixels along 
 the horizontal (X-axis) and vertical (Y-axis) directions as two possible operations. We also 
 define the rotation operations of $60^\circ$ and $120^\circ$ for a given image. 
 
 So, 
 from original image location of $(x_c,y_c)$ we can have 
 combinations of $(x_c\pm k\Delta, y_c\pm k\Delta)$ where $k\in \{0,1,2\}$. These
 translation operations will give us $25=(5\times 5)$ times the original 
 images. The two rotation operations will give us 3 times the images. 
% After getting $75(=5\times 5\times 3)$ times the original set of images, we can use the operations of 
% flip and rotation by
% $90^\circ$, $180^\circ$, and $270^\circ$ to augment the dataset by 8 times.
 Eventually, one original image will give us $75(=5\times 5\times 3)$ images.

The originally extracted images were augmented using the method described above. The images were then checked manually 
for areas with empty regions which appear due to rotation and translation into empty WSI area. We have removed these images
and then sorted all the remaining 
images which we extracted according to patient and label. 
 There are $53557$ `Cribriform'
 and $110151$ `Non-cribriform' images after augmentation. 
 This way we have a total of $163708$ images $(1024\times 1024$ pixels) from $19$ TCGA patients.
 Table~\ref{tab:cribriform_augmented_dataset} tabulates the patient wise number of images in the
 augmented dataset.

\begin{table}[htbp]
\centering
	\caption{Description of the all 	images in the \textbf{augmented} cribriform dataset. 
	We have 12 unique cribriform and 7 unique 
	non-cribriform patients. These images are of $1024 \times 1024$ pixels.
	}
	\label{tab:cribriform_augmented_dataset}

\begin{tabular}{|p{0.8cm}|p{3.2cm}|p{3.5cm}|p{2.5cm}|p{2cm}|} \hline
\textbf{S.N.}&\textbf{Slide Name \quad (Patient ID)} &\textbf{Gleason grade} &\textbf{Number of Cribriform Images} &\textbf{Number of Non-cribriform Images} \tabularnewline \hline
1 & TCGA-2A-A8VO & 3+3 (HG-PIN) & - & 1292 \\ \hline
2 & TCGA-2A-A8VT & 3+3 (HG-PIN) & 152 & -\\ \hline
3 & TCGA-EJ-5510 & 4+3 (HG-PIN) & 456 & 76\\ \hline
4 & TCGA-EJ-5511 & 3+4 (HG-PIN) & 76 & 1216\\ \hline
5 & TCGA-EJ-5519 & 4+4 (HG-PIN) & 380 & -\\ \hline
6 & TCGA-EJ-7791 & No report & 76& 21201 \\ \hline
7 & TCGA-EJ-7797 & 3+4 (HG-PIN) & - & 1596\\ \hline
8 & TCGA-EJ-8469 & 4+5 (HG-PIN) & 24000 & - \\ \hline
9 & TCGA-EJ-A46F & 4+4 (HG-PIN) & 10594 & - \\ \hline
10 & TCGA-FC-7708 & No report & 379 & 29935 \\ \hline

11 & TCGA-G9-6338 & 4+3 (No HG-PIN) & - & 2736\\ \hline
12 & TCGA-G9-6363 & 4+3 (HG-PIN) & - & 1064\\ \hline
13 & TCGA-HC-7078 & No report & 20 & 5188 \\ \hline
14 & TCGA-HC-7211 & 3+4 (HG-PIN) & 1900 & -\\ \hline
15 & TCGA-HC-7212 & 3+4 (HG-PIN) & 1292 & -\\ \hline
16 & TCGA-HC-7820 & 3+4 (HG-PIN) & - & 3943 \\ \hline
17 & TCGA-XJ-A9DI & 5+4 (No HG-PIN) & - & 11699 \\ \hline
18 & TCGA-XK-AAJP & 4+3 (HG-PIN) & -& 30185 \\ \hline
19 & TCGA-YL-A8HL & 4+5 (No HG-PIN) & 14233 & - \\ \hline
\multicolumn{3}{|c|}{\textbf{Total (163708 images from 19 patients)}} & \textbf{53557} & \textbf{110151} \\ \hline
\end{tabular}
\end{table}

As the total number of images in the augmented dataset  is quite big, we used a subset of 
images for our experiments. We have 
defined three sets of patients for a three-fold cross-validated study such that 
patients for training, validation, and testing images are mutually exclusive. This configuration
 is to mimic the real world scenario for deployment of any cribriform pattern classification
system.
Table~\ref{tab:patient_list} tabulates these sets along with 
their use during the three folds. 	We sampled $1500$ Cribriform 
(+ve), $1500$ Non-cribriform (-ve) images in 
each of these sets for use in our experiments.
This way we have a balanced dataset in our studies. 
We 
also defined an additional unseen test set for further evaluating our models. 
This additional unseen test set 
contains the images which have never been used for training, validation, and testing in the three-fold cross-validated 
study. The patients in the additional unseen test set and the test set in the cross-validated study for a given fold are the same. 
The addition unseen test set also contains $1500$ Cribriform (+ve) and $1500$ 
Non-cribriform (-ve) images in each fold ( three folds, same as the cross-validated study). 
%We expect our models to perform similarly during testing in both of the cross-validated study and the additional unseen test set. 

\begin{table}[htbp]
\centering
	\caption{Set of patients in the three-fold cross-validated study. 
	We sampled $1500$ Cribriform (+ve), $1500$ Non-cribriform (-ve) images in 
	each of these sets for use in our experiments.}
	\label{tab:patient_list}

\begin{tabular}{|p{4.1cm}|p{4.1cm}|p{4.1cm}|} \hline
\textbf{Set 1}&
\textbf{Set 2}&
\textbf{Set 3} \tabularnewline \hline

\textbf{Fold 01: Train;} &
\textbf{Fold 01: Validation;}&
\textbf{Fold 01: Test;} \tabularnewline \hline

\textbf{Fold 02: Validation;}&
\textbf{Fold 02: Test;} &
\textbf{Fold 02: Train;} \tabularnewline \hline

\textbf{Fold 03: Test;}&
\textbf{Fold 03: Train;} &
\textbf{Fold 03: Validation;} \tabularnewline \hline

\begin{itemize}
\item TCGA-2A-A8VT, 
\item TCGA-HC-7212,
\item TCGA-FC-7078,
\item TCGA-YL-A8HL,
\item TCGA-XJ-A9DI,
\item TCGA-XK-AAJP. 
\end{itemize}
$16056$ Cribriform (+ve),
$71839$ Non-cribriform (-ve) 
& 
\begin{itemize}
\item TCGA-2A-A8VO,
\item TCGA-EJ-7791,
\item TCGA-EJ-7797, 
\item TCGA-HC-7211,
\item TCGA-EJ-5519,
\item TCGA-G9-6363,
\item TCGA-EJ-A46F.

\end{itemize}
$12949$ Cribriform (+ve),
$25153$ Non-cribriform (-ve) 
& 
\begin{itemize}
\item TCGA-HC-7708,
\item TCGA-HC-7820,
\item TCGA-EJ-5510,
\item TCGA-G9-6338,
\item TCGA-EJ-5511,
\item TCGA-EJ-8469.
\end{itemize}
$24552$ Cribriform (+ve),
$13159$ Non-cribriform (-ve) 

\\ \hline
$1500$ Cribriform (+ve),
$1500$ Non-cribriform (-ve) &
$1500$ Cribriform (+ve),
$1500$ Non-cribriform (-ve) &
$1500$ Cribriform (+ve),
$1500$ Non-cribriform (-ve) \\ \hline
\end{tabular}
\end{table}

\section{Methods}

%Describe the overall idea of using different methods and so on. 
%One paragraph should tell the essence of the methods. 
We have studied nuclei feature based classical machine learning model along with 
fine-tuned deep learning models for cribriform pattern detection. The classical machine learning 
model act as a base-line method for our system. We discuss all the methods for 
cribriform pattern detection in following sections. 

\subsection{Nuclei features with SVM}
\label{sec:svm_cellprofiler}

Various image based automated PCa
grading studies have suggested using local and global
features derived from nuclei patterns~\cite{kwak2017nuclear,fukuma2016study,khan2017predicting,ali2013cell}.
Most commonly used local features quantify 
intensity distribution, 
radial intensity distribution, etc inside the segmented nuclei objects. 
These studies have also suggested creating nuclei graphs to quantify nuclei spatial
distribution as a global feature. These nuclei based features with SVM are used as a base-line method for 
cribriform pattern detection experiments. 
%For global features 
%quantifying nuclei spatial distribution, these 
%studies have suggested creating nuclei-graphs and extracting features from it. 

%Fukuma~et al.~\cite{fukuma2016study} used CellProfiler~\cite{carpenter2006cellprofiler}
% for nuclei segmentation.
Given a nuclei segmentation, 
a digraph $\mathcal{G}$ can be defined whose vertices are the centroids of the segmented nuclei~\cite{fukuma2016study}. 
 $\mathcal{G}$ is a complete 
digraph with edges weighted according to euclidean distance between the vertices (centroids). 
 The nuclei spatial distribution was then quantified by computing
 Delaunay Triangulation and Minimum Spanning Tree (MST). 
The Delaunay Triangulation for the vertices in $\mathcal{G}$ was
computed using the Triangle software~\cite{shewchuk1996triangle}. %provided by Jonathan Shewchuk @ UC Berkeley
 The triangle area and perimeter based sub-features are extracted from this Delaunay Triangulation.
The MST for $\mathcal{G}$ was also computed using Kruskal's algorithm~\cite{kruskal1956shortest}.
For a given MST, its corresponding edge weight distribution was quantified as a sub-feature.
Both of these sub-features constitute the image level nuclei feature.

\begin{figure}[htbp]
\begin{center}
\fbox{\includegraphics[width=0.5\textwidth]{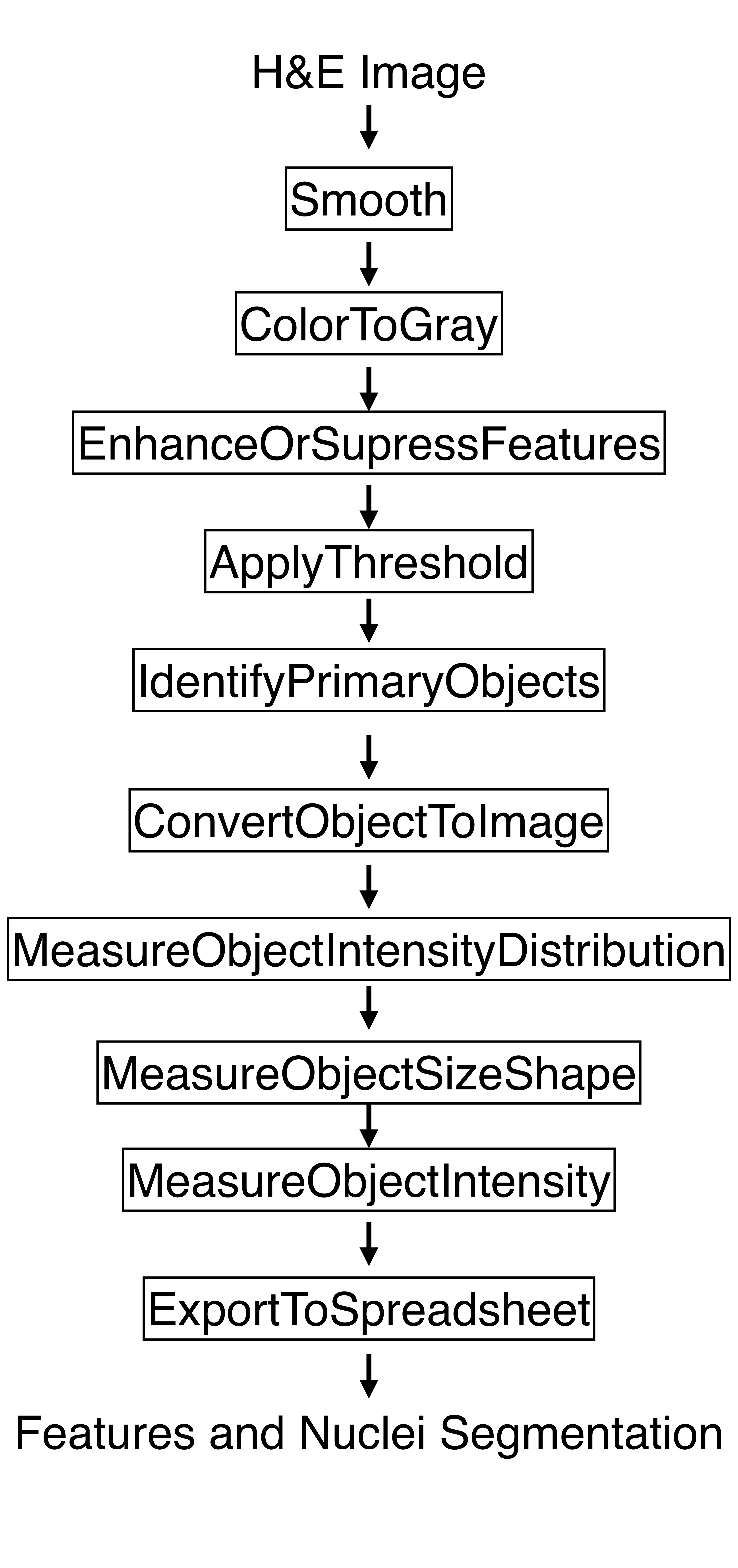}}
\caption{Modules used in the CellProfiler~\cite{carpenter2006cellprofiler} pipeline for nuclei segmentation. 
This pipeline has been implemented as proposed by Fukuma~et al.~\cite{fukuma2016study}.}
\label{fig:cellprofiler_pipeline}
\end{center}
\end{figure}

\begin{figure}[htbp]
\begin{center}
\includegraphics[width=\textwidth]{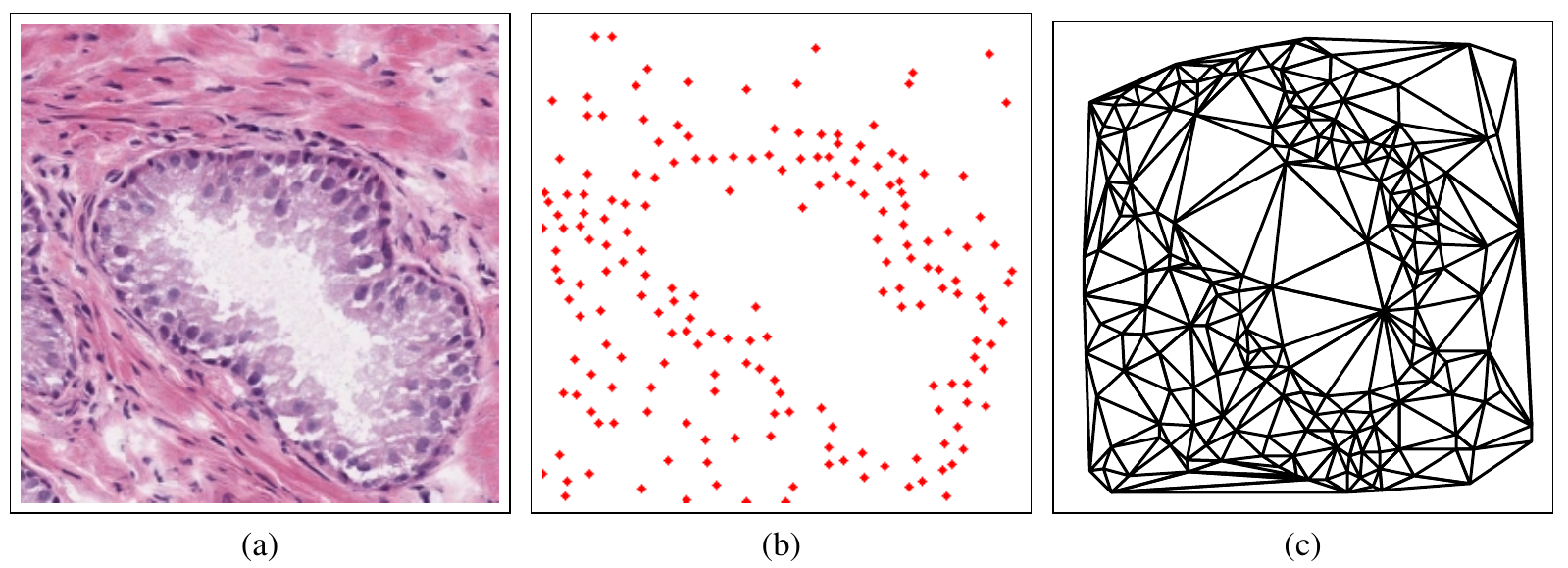}
\caption{Intermediate stages during nuclei feature generation for an input H\&E image using  
CellProfiler~\cite{carpenter2006cellprofiler} 
and Delaunay Triangulation. 
(a) Example input H\&E image. (b) Segmented nuclei locations are indicated in red diamonds (By 
CellProfiler~\cite{carpenter2006cellprofiler}). Graph $\mathcal{G}$ is defined using these nuclei locations. (c) Delaunay Triangulation 
using the vertices of graph $\mathcal{G}$. Table~\ref{tab:PCa_features} discusses these features in detail.}
\label{fig:steps}

\end{center}
\end{figure}

The CellProfiler~\cite{carpenter2006cellprofiler} pipeline 
suggested by Fukuma~et al.~\cite{fukuma2016study} has been used for nuclei segmentation 
and feature extraction.
% CellProfiler~\cite{carpenter2006cellprofiler} can provide us nuclei level features like intensity distribution, 
%radial intensity distribution,~\emph{etc}. after nuclei segmentation for a given H\&E image. 
 Fig.~\ref{fig:cellprofiler_pipeline} shows the modules used in the CellProfiler~\cite{carpenter2006cellprofiler} pipeline.
 %This pipeline has been implemented according to Fukuma~et al.~\cite{fukuma2016study}.
Fig.~\ref{fig:steps}(a) shows a sample input H\&E image for the 
CellProfiler~\cite{carpenter2006cellprofiler} pipeline.
Fig.~\ref{fig:steps}(b) shows the segmented nuclei locations as red diamonds on white background.
These nuclei locations are used to define $\mathcal{G}$. 
These segmented nuclei regions 
 are also used to extract nuclei level features like intensity distribution, eccentricity, etc. 
 The MST features are extracted using the vertices in $\mathcal{G}$. 
Fig.~\ref{fig:steps}(c) shows the Delaunay Triangulation using the vertices in $\mathcal{G}$.
% The triangle area and perimeter based features are extracted using this Delaunay Triangulation.
 Table~\ref{tab:PCa_features} discusses these features in detail. This table also details which 
tool or algorithm or CellProfiler~\cite{carpenter2006cellprofiler} module was used for the given 
nuclei sub-feature extraction.

 Kwak~et al.~\cite{kwak2017nuclear} illustrated that the RBF kernel SVM performs 
 better then polynomial kernel SVM for the above nuclei features. Following this idea,
 the $C$ and $\gamma$ for the RBF kernel were fine-tuned first and then fixed as $C=100$ and $\gamma=0.1$ for final experiments.

%
%The idea for base-line experiments has two main steps:
%\begin{enumerate}
%\item \textbf{Feature extraction}: using tools like CellProfiler~\cite{carpenter2006cellprofiler} and in-house C++ implementations. 
%\begin{enumerate}
%\item Nuclei segmentation using CellProfiler~\cite{carpenter2006cellprofiler}. 
%
%\item Compute features as illustrated in Table~\ref{tab:PCa_features} using 
%CellProfiler~\cite{carpenter2006cellprofiler} and available C++ code. 
%Fig.~\ref{fig:steps}(a) shows a sample input image for the CellProfiler Pipeline.
%Fig.~\ref{fig:steps}(b) shows the segmented nuclei location as red diamonds on white background.
% The MST features are extracted using the nuclei locations in this image. These segmented nuclei regions 
% are also used to extract nuclei level features like intensity distribution, eccentricity,~\emph{etc}. 
%Fig.~\ref{fig:steps}(c) shows the Delaunay Triangulation using nuclei centroids as vertices.
% The triangle area and perimeter based. features are extracted using this Delaunay Triangulation.
%
%
%\end{enumerate}
%
%\item \textbf{Classification}: Using SVM. 

%\end{enumerate}

%
%
%
%\begin{figure}[htbp]
%\centering
%\subfloat[][]{\fbox{\includegraphics[width=0.30\textwidth]{./figures/TCGA-EJ-7797_L47084_T29335_W1024_H1024_W256_H256.jpg}}}$~$
%\subfloat[][]{\fbox{\includegraphics[width=0.30\textwidth]{./figures/TCGA-EJ-7797_L47084_T29335_W1024_H1024_W256_H256_mask.png}}}$~$
%\subfloat[][]{\fbox{\includegraphics[width=0.30\textwidth]{./figures/TCGA-EJ-7797_L47084_T29335_W1024_H1024_W256_H256.pdf}}}
%\caption{(a) Example input H\&E image. (b) Segmented nuclei locations are indicated in red diamonds. Graph 
%$\mathcal{G}$ is defined using these nuclei locations.
% (c) Delaunay Triangulation using the vertices of graph $\mathcal{G}$.}
%\label{fig:steps}
%\end{figure}
%
%

\begin{landscape}

%
%\begin{table}
%	\caption{Nuclei features for Cribriform Pattern Detection}
%\begin{tabular}
%{|p{7cm}|p{4cm}|p{4cm}|p{7cm}|}\hline
%\label{tab:PCa_features}
%

\begin{longtable}{|p{7cm}|p{4cm}|p{4cm}|p{7cm}|}
	\caption{Nuclei features for cribriform pattern detection}\tabularnewline \hline
\label{tab:PCa_features}
 \textbf{Feature (Total Dimensions: 57)} &
 \textbf{CellProfiler Module / Tool} &
 \textbf{Module and Feature Description} &
 \textbf{Relevance of feature with respect to PCa Histopathology}\tabularnewline \hline
\endfirsthead %lines to appear as the head of table in first page
\multicolumn{4}{c}{ \tablename\quad \thetable \quad \textit{Continued from previous page}} \tabularnewline \hline
 \textbf{Feature (Total Dimensions: 57)} &
 \textbf{CellProfiler Module / Tool} &
 \textbf{Module and Feature Description} &
 \textbf{Relevance of feature with respect to PCa Histopathology}\tabularnewline \hline
 \endhead %lines to appear as the head of table in every page except the first one
\hline \multicolumn{4}{r}{\textit{Continued on next page}}\tabularnewline 
\endfoot %lines to appear as the foot of table in every page except last
\hline
\endlastfoot%lines to appear as the head of table in last page

%
%
% \textbf{Feature (Total Dimensions: 57)} &
% \textbf{CellProfiler Module / Tool} &
% \textbf{Module and Feature Description} &
% \textbf{Relevance of feature with respect to PCa Histopathology} \tabularnewline \hline 

Number and area of nuclei \cite{fukuma2016study, kwak2017nuclear} 
The average ($\mu$), standard deviation ($\sigma$),
disorder $(1-\frac{1}{1+{\frac{\mu}{\sigma}}})$, and minimum to maximum ratio 
of area is computed.

\textbf{Dimensions:~5}

& 
MeasureImageArea and IdentifyPrimaryObjects. & 
Measures the area and number of a given nuclei in the image. & 
%Nuclei density and size are important in PCa assessment. \tabularnewline \hline 
%\multirow{3}{7cm}{The morphology, size, and intensity distribution 
%of nuclei are important in PCa assessment.}
%\tabularnewline \cline{1-3}
The morphology, size, and intensity distribution of nuclei are important in PCa assessment. \tabularnewline \hline

Radial Distribution of Pixel Intensity of the nuclei \cite{fukuma2016study,kwak2017nuclear} 
Example features: Mean Intensity and Mean Intensities along the four rings(bins).
The average ($\mu$), standard deviation ($\sigma$),
disorder $(1-\frac{1}{1+{\frac{\mu}{\sigma}}})$, and minimum to maximum ratio 
of these two measurements are computed.

\textbf{Dimensions:~20}& 
MeasureObjectIntensity Distribution and
MeasureObjectIntensity &
Given an image with objects (nuclei) identified, these
modules measures the intensity distribution from each object's center 
to its boundary within a user-controlled number of bins, i.e. rings. 
& The morphology, size, and intensity distribution of nuclei are important in PCa assessment. \tabularnewline \hline
%\tabularnewline \cline{1-3}

Nucleus Size and Shape \cite{fukuma2016study,kwak2017nuclear}. 
The nuclei shape can be modelled as an ellipse and subsequent features will be 1) minor axis length, 2) major axis length, 
3) eccentricity, 4) orientation, 5) solidity.

The average ($\mu$), standard deviation ($\sigma$),
disorder $(1-\frac{1}{1+{\frac{\mu}{\sigma}}})$, and minimum to maximum ratio 
of these five measurements are computed.

\textbf{Dimensions:~20} & 
MeasureObjectSizeShape &
Given an image with identified objects (e.g. nuclei or cells), this module extracts individual area and shape feature. 
& The morphology, size, and intensity distribution of nuclei are important in PCa assessment. \tabularnewline \hline
%\tabularnewline \hline \hline 

Minimum Spanning Tree (MST)\cite{fukuma2016study,kwak2017nuclear}
The edge weights are computed as the distance between nuclei centroids. 
The average ($\mu$), standard deviation ($\sigma$), disorder $(1-\frac{1}{1+{\frac{\mu}{\sigma}}})$, and minimum to maximum ratio of the edge weights are features.

\textbf{Dimensions:~4}
 & Kruskal's algorithm~\cite{kruskal1956shortest}. & 
A MST is created using the nuclei centroids. Khan~et al.~\cite{khan2017predicting} 
however, mentions that just MST alone does not generate enough 
features to differentiate between images with Cribriform pattern (Gleason 4) 
from images with Gleason pattern 3. & 
These features quantify the information specific to the spatial distribution of nuclei in the given field of view. 
The nuclei spatial distribution provides image level information which is important in PCa assessment. 
Khan~et al.~\cite{khan2017predicting} provides additional insights about MST. The mean edge length 
of MST characterises the degree to which the epithelial nuclei are invading the stroma surrounding the gland.
%The features of degree distribution provide tumor architecture insights. 
%Higher proportions of degree 1 vertices are expected to correspond to higher Gleason grades,
%whereas higher proportions of degree 2 vertices are expected to correlate negatively with the Gleason grade.
%\tabularnewline \cline{1-3}

\tabularnewline \hline

Delaunay Triangulation \cite{fukuma2016study}.
The area and perimeter of each triangle is computed, and the 
 average ($\mu$), standard deviation ($\sigma$),
disorder $(1-\frac{1}{1+{\frac{\mu}{\sigma}}})$, and minimum to maximum ratio 
of area and perimeter are computed.

\textbf{Dimensions:~8}
& Triangle~\cite{shewchuk1996triangle}. &
A Delaunay Triangulation is created using the nuclei centroids. & 

These features quantify the information specific to the spatial distribution of nuclei in the given field of view. 
The nuclei spatial distribution provides image level information which is important in PCa assessment. 
Khan~et al.~\cite{khan2017predicting} provides additional insights about MST. The mean edge length 
of MST characterises the degree to which the epithelial nuclei are invading the stroma surrounding the gland.

\tabularnewline \hline 

\end{longtable}
%\end{tabular}
%\end{table}

\end{landscape}

%\subsection{fine-tuning of `ResNet-50', `ResNet-101', `DenseNet-121', `DenseNet-169', and `Inception-v3'}
\subsection{Fine-tuning of pre-trained DL architectures}
\label{sec:fine_tune_ResNet}

%Once we have the extracted images in $1024 \times 1024$ pixels dimensions we performed 
%our experiments with fine-tuning of different state-of-art DL architectures
%which has been pre-trained on the ImageNet dataset~\cite{russakovsky2015imagenet}. 

The extracted images at $1024 \times 1024$ pixels dimensions were used for 
experiments with fine-tuning of different state-of-art DL architectures. These state-of-art 
DL architectures have been pre-trained on the ImageNet dataset~\cite{russakovsky2015imagenet}. 
 Fine-tuning was done in two stages as follows:
 \begin{enumerate}
 \item The last layers of each pre-trained network were 
 modified for the cribriform pattern classification (binary classification). 
 All the layers except the last fully connected layers in the modified network were frozen (non-trainable) for the first stage.
 The modified network was trained for 100 epochs. 
 
 \item In the second stage the last block before the fully connected layers in the modified network was set as trainable.
 In this second stage, the last block along with the fully connected layers were trained for 100 epochs. 
 
 \end{enumerate}
 
 For both of the above stages, the learning rate was kept low to prevent overfitting due large number of trainable parameters with the given small 
 amount of training images. The two stage fine-tuning strategy has been borrowed from the online Keras~\cite{chollet2015keras} 
 tutorial ``Building powerful image classification models using very little data''~\cite{keras_tune}. This tutorial used
 TensorFlow~\cite{abadi2015tensorflow} as a back-end for deep learning.

Another possible strategy for fine-tuning is skipping first stage and directly fine-tune at the 
second stage itself. This way one will get non-reliable 
results because the random initialisation (high entropy) of last fully connected layers 
will induce massive change in weights in the last block of the network. The first stage in the above used strategy essentially 
reduces the entropy in the last fully connected layer leading to reliable results. 
%
%\vspace{1cm}
%{\color{red} \textbf{Mundher}: Please add your method's description here. We will also update the Table~\ref{tab:results} accordingly.}

\subsection{Fine-tuning of pre-trained and modified ResNet architectures}

Additionally, we fine-tuned ResNet-50~\cite{he2016ResNet,he2016identity} and ResNet-22, whereby we replaced the 
output layer of ResNet with two output nodes and kept all previous layers untouched. 
We separated the whole fine-tuning procedure into two stages. In the first stage, only 
the last layer was fine-tuned which runs for $\frac{1}{3}^\text{rd}$ of the total number of epochs. 
For the second stage which runs for $\frac{2}{3}^\text{rd}$ of the total number of epochs, the last 
ResNet block was trained as well as the output layer. In ResNet, all blocks are a 
bottle-neck block that consists of 3 convolutional layers. 

ResNet-22 is a modified version of ResNet-50~\cite{he2016ResNet} whose structure is basically the first 
21 layers of ResNet-50~\cite{he2016ResNet} plus a fully-connected layer at the output. 
The main advantage of using ResNet-22 is that it has a fewer number 
of parameters while still maintaining the powerful capabilities of the original ResNet~\cite{he2016ResNet} 
architecture. The input size is $256\times 256$, whereby each image has three channels, 
namely R, G, and B. Architecture comparisons between the ResNet-50 and ResNet-22 network 
architecture are tabulated in Table~\ref{tab:ResNet50_22}. 
Both models share the same architecture for the 
first 21 layers as shown in Table~\ref{tab:ResNet50_22} at the first four rows.

\begin{table}
	\caption{
	 Comparison between the network architectures of ResNet-50~\cite{he2016ResNet} and ResNet-22. 
	 Each [] means one residual block. For example, in the fourth row we used 4 residual blocks where each 
	 residual block consists of 1
	 $\times$1 convolution followed by $3\times 3$ and then $1\times 1$. Because ResNet-22 duplicates only the 
	 first 21 layers of ResNet-50~\cite{he2016ResNet}, the sixth and the seventh row has `No Operation'.}

	\label{tab:ResNet50_22}
\centering
\begin{tabular}{|p{3cm}|p{4cm}|p{4cm}| }\hline
\textbf{Output Size} &\textbf{ResNet-50~\cite{he2016ResNet}} &\textbf{ResNet-22} \tabularnewline \hline

262 $\times$ 262 & \multicolumn{2}{c|}{ 1$\times$1, 64, stride 2} \\ \hline
63 $\times$ 63 & \multicolumn{2}{c|}{ 3$\times$3, Max-Pool , stride 2} \\ \hline
63 $\times$ 63 & \multicolumn{2}{c|}{ 

~$\begin{bmatrix}
1\times 1, 64 \\
3\times 3, 64 \\
1 \times 1, 256 \end{bmatrix} \times 3
 $~
} \\ \hline

16 $\times$ 16 & \multicolumn{2}{c|}{ 

~$\begin{bmatrix}
1\times 1, 128 \\
3\times 3, 128 \\
1 \times 1, 512 \end{bmatrix} \times 4
 $~
} \\ \hline

16 $\times$ 16 & 

~$\begin{bmatrix}
1\times 1, 256 \\
3\times 3, 256 \\
1 \times 1, 1024 \end{bmatrix} \times 6
 $~& No operation
 \\ \hline

8 $\times$ 8 & 
~$\begin{bmatrix}
1\times 1, 512 \\
3\times 3, 512 \\
1 \times 1, 2048 \end{bmatrix} \times 3
 $~& No operation
 \\ \hline

1 $\times$ 1 & 
\multicolumn{2}{c|}{ Average Pool, 2-D, Full-Connected, Softmax} \\ \hline

\end{tabular}
\end{table}

\subsection{Feature combination using Multi-Layer Perceptron (MLP)}

Kallen~et al.~\cite{kallen2016towards} proposed using OverFeat~\cite{sermanet2013overfeat} network for 
feature extraction from prostate H\&E images. These features were then fed into an SVM for automated PCa grading. 
During the experiments with nuclei features and various deep learning models 
 some scope of improvement for cribriform pattern detection was observed. 
Subsequently, these methods were combined using feature concatenation and training a Multi-Layer Perceptron (MLP).
Following a similar approach to Kallen~et al.~\cite{kallen2016towards} features from a given image 
were extracted and then used to fine-tune the pre-trained `ResNet', `DenseNet', and `Inception-v3' models.

In the MLP, the 57 nuclei 
features are concatenated with features from all 
``VGG',`ResNet', `DenseNet', and `Inception-v3' models trained upon 
$256 \times 256$,
 $128 \times 128$, $64 \times 64$, $32 \times 32$, and $16 \times 16$ pixels images. The two 
hidden layers in this MLP has $512$ and $128$ nodes respectively. This MLP was trained for $10,000$ epochs. We achieved testing 
accuracy of $85.93~\pm~7.54$ across three folds.

\section{Results}
\label{sec:results}

Several DL models and nuclei features based model were assessed for effectiveness using the
 augmented cribriform image (balanced) dataset. 
The H\&E images in the dataset were downscaled to $256 \times 256$,
 $128 \times 128$, $64 \times 64$, $32 \times 32$, and $16 \times 16$
% $8 \times 8$, and $4 \times 4$
 pixels for fine-tuning and testing of all DL models. The nuclei feature based SVM was trained and evaluated with images downscaled 
 to $256 \times 256$ pixels. 
 The Keras~\cite{chollet2015keras} based framework resizes the input images to the internal 
image dimension of the given DL network. For example, `ResNet-50' uses `$224 \times 224$' pixels 
input image resolution. Given an input image of $256 \times 256$ pixels, it is resized to 
$224 \times 224$ pixels and then fed into the network during training and testing. The 
same process is used for all the DL models with different input image sizes (scales).

Three-fold cross-validated study was done such that 
patients for training, validation, and testing images are mutually exclusive 
to mimic the real world scenario for a cribriform pattern classification system. 
As discussed before in section~\ref{sec:data_augmentation}, we had also defined 
an additional unseen test set for further evaluating our models. 
We expect the trained models to perform similarly during testing in both of the 
cross-validated study and the additional unseen test set. 
We tested the top three performing 
individual DL models on this additional unseen test set across the three folds. 

Table~\ref{tab:results} tabulates the testing accuracy for nuclei feature 
based methods along with fine-tuned DL architectures in the three-fold cross-validated study and for the 
top three models on the additional 
unseen test sets. The results for the top three models in the three-fold cross-validation study and
on the additional unseen test set were similar. 
 The experiments were conducted in two separate locations. The nuclei feature 
based method along with fine-tuned DL architectures were evaluated at first location. The modified ResNet~\cite{he2016ResNet}
was designed and implemented in second location. The implementations were shared across the locations to validate reproducibility. 
For reproducibility checks, the DL experiments were done using 300 images on both locations. The results on both locations were identical. 
First experiment location used Ubuntu 14.04 64bit desktop with 32GB RAM, Intel i7 3.5 GHz CPU, and 6GB Nvidia TITAN GPU. 
The second location used Ubuntu 16.04 64bit desktop with 64 GB RAM, Intel i7 3.4 GHz CPU, and 12 GB Nvidia Titan X GPU.

% First experiment location used Ubuntu 14.04 64bit desktop with 32GB RAM and 6GB Nvidia TITAN GPU. The second location 
%used XXXX $\ldots$.
%
%
%{\color{red} Mundher please give me your machine/cluster's specification. I will add here.}
%The best testing results were achieved when all the methods trained upon at different scales were combined together by MLP.

%{\color{red} ------ Discussion of performance variance w.r.t scale in various DL model goes here. TO BE UPDATED.. ------}

%Fig.~\ref{fig:result_discussion} shows one `Cribriform' and one `Non-cribriform' image at different scales. 
%This figure also illustrated the example images at the resolution of $256 \times 256$, 
%$128 \times 128$, $64 \times 64$, $32 \times 32$, and $16\times 16$ pixels along with their corresponding resized $224\times 224$
% pixels resolution images. As discussed earlier, 
% the H\&E images were originally extracted as $1024 \times 1024$ pixels from the whole slide image and then 
% downscaled to the resolutions of $256 \times 256$, $128 \times 128$, $64 \times 64$, $32 \times 32$, and $16\times 16$ pixels.
%The DL models in the experiments viz. `ResNet-50', `ResNet-101', 'DenseNet-121', `DenseNet-169' used the image 
%resolution of $224 \times 224$ pixels and the images are resized to this resolution during training and testing.
%Similarly, `Inception-v3' uses $299 \times 299$ pixels resolution and our input images at multiple scales at resized for
%training and testing purposes.

\subsection{Performance of DL models}

Given the images rescaled to different resolutions from same image, the amount of usable information 
is directly proportional to the resolution of the rescaled image. We studied the performance from 
$256\times 256 $ to $16\times 16$ pixels, the test accuracy decreases with image resolution  which is 
as per our  expectations. 

VGG16~\cite{simonyan2014very}, VGG19~\cite{simonyan2014very}, and Inception-v3~\cite{szegedy2015going,szegedy2016rethinking} 
 were the top performers while newer and more complex architectures 
 ResNet-50~\cite{he2016ResNet}, DenseNet-121~\cite{huang2017densely}, and DenseNet-169~\cite{huang2017densely} did not perform well. 
This indicates that DL architectures, with low number trainable parameters (low model complexity) performed better than 
the DL architectures with much higher number of trainable parameters(high model complexity). This results can be attributed to the 
fact that the highly complex DL architectures will need higher number of training data samples. The same results were observed 
when ResNet-22 was designed after modifying ResNet50~\cite{he2016ResNet}. 

The additional unseen test set results for our top three performing models 
VGG16~\cite{simonyan2014very}, VGG19~\cite{simonyan2014very}, and 
Inception-v3~\cite{szegedy2015going,szegedy2016rethinking} were similar to the three-fold cross-validated study results. This further confirms their 
robust performance. Also, in some of our trained/fine-tuned models, we observed 
that standard error of testing accuracy is a bit high indicating variable model 
performance across three folds. This can be attributed to the 
low number of patients being used for training. The modelsÕ performance 
will improve with more patient information.

\begin{table}
	\caption{Testing Accuracy for various methods.
	Reported values are \textbf{average} $\pm$ \textbf{standard error} across the three folds.
%{\color{red} Results for grayscale images will also be added in similar fashion}
	VGG16~\cite{simonyan2014very}, VGG19~\cite{simonyan2014very}, Inception-v3~\cite{szegedy2015going,szegedy2016rethinking}, and combination of all DL methods along with nuclei features using MLP
	achieve best results (Indicated in bold).
	}
	\label{tab:results}
\centering
\begin{tabular}{|p{6cm}|p{5cm}|p{5cm}| }\hline
\textbf{Method} &\textbf{Input image dimensions (RGB), Scale} &\textbf{Testing Accuracy ($\%$age), Testing accuracy on additional unseen set (if applicable in $\%$age) } \tabularnewline \hline
%
%RBF kernel SVM ($C=100$, $\gamma=0.1$) 
%using nuclei features. %(described in Table~	\ref{tab:PCa_features})
%&$256 \times 256$ pixels, 1:1 & 0.76 $\pm$ 0.05 & 1.00 $\pm$ 0.01\\ \hline \hline

`ResNet-22': $256\times 256$ pixels. & $256 \times 256$ pixels, 1:1 & 73.33 $\pm$ 16.66 \\ \hline \hline

\multirow{5}{7cm}{\pbox{7cm}{`VGG16'~\cite{simonyan2014very}: \\ $224 \times 224$ pixels.}} &$256 \times 256$ pixels, 1:1 & \textbf{85.65} $\pm $ \textbf{6.68}, \textbf{ 85.81} $\pm $ \textbf{6.74} \\ \cline{2-3}

 &$128 \times 128$ pixels, 1:2 & $81.08 \pm 4.58 $ \\ \cline{2-3}
							 &$64 \times 64$ pixels, 1:4 & $72.33 \pm 6.19 $ \\ \cline{2-3}
 							 &$32 \times 32$ pixels, 1:8 & $56.08 \pm 7.23 $ \\ \cline{2-3}
							&$16 \times 16$ pixels, 1:16 & $75.46 \pm 7.73 $ \\ \hline \hline

\multirow{5}{7cm}{\pbox{7cm}{`VGG19'~\cite{simonyan2014very}: \\ $224 \times 224$ pixels.}} &$256 \times 256$ pixels, 1:1 & \textbf{86.78} $\pm$ \textbf{6.97}, \textbf{ 86.25} $\pm $ \textbf{7.18} \\ \cline{2-3}
 &$128 \times 128$ pixels, 1:2 & $83.76 \pm 9.47 $ \\ \cline{2-3}
							 &$64 \times 64$ pixels, 1:4 & $81.10 \pm 7.54 $ \\ \cline{2-3}
 							 &$32 \times 32$ pixels, 1:8 & $50.14 \pm 0.18 $ \\ \cline{2-3}
							&$16 \times 16$ pixels, 1:16 & $73.92 \pm 11.28 $ \\ \hline \hline

\multirow{5}{7cm}{\pbox{7cm}{`Inception-v3'~\cite{szegedy2015going,szegedy2016rethinking}: \\ $299 \times 299$ pixels.}} &$256 \times 256$ pixels, 1:1 & \textbf{88.18} $\pm$ \textbf{5.99},\textbf{ 88.04} $\pm $ \textbf{5.63}\\ \cline{2-3}
 &$128 \times 128$ pixels, 1:2 & $84.37 \pm 8.22 $ \\ \cline{2-3}
							 &$64 \times 64$ pixels, 1:4 & $82.37 \pm 9.78 $ \\ \cline{2-3}
 							 &$32 \times 32$ pixels, 1:8 & $79.83 \pm 8.88 $ \\ \cline{2-3}
							&$16 \times 16$ pixels, 1:16 & $80.84 \pm 9.38 $ \\ \hline \hline

\multirow{5}{7cm}{\pbox{7cm}{`DenseNet-121'~\cite{huang2017densely}: \\ $224 \times 224$ pixels.}} &$256 \times 256$ pixels, 1:1 & $73.48 \pm 9.76 $ \\ \cline{2-3}
 &$128 \times 128$ pixels, 1:2 & $65.20 \pm 10.8$ \\ \cline{2-3}
							 &$64 \times 64$ pixels, 1:4 & $63.02 \pm 7.93 $ \\ \cline{2-3}
 							 &$32 \times 32$ pixels, 1:8 & $63.74 \pm 13.86$ \\ \cline{2-3}
							&$16 \times 16$ pixels, 1:16 & $59.64 \pm 10.75 $ \\ \hline \hline

\multirow{5}{7cm}{\pbox{7cm}{`DenseNet-169'~\cite{huang2017densely}: \\ $224 \times 224$ pixels.}} &$256 \times 256$ pixels, 1:1 & $64.91 \pm 6.33 $ \\ \cline{2-3}
 &$128 \times 128$ pixels, 1:2 & $67.12 \pm 11.26 $ \\ \cline{2-3}
							 &$64 \times 64$ pixels, 1:4 & $61.65 \pm 7.45 $ \\ \cline{2-3}
 							 &$32 \times 32$ pixels, 1:8 & $54.67 \pm 3.01 $ \\ \cline{2-3}
							&$16 \times 16$ pixels, 1:16 & $56.78 \pm 4.23 $ \\ \hline \hline

\multirow{5}{7cm}{\pbox{7cm}{`ResNet-50'~\cite{he2016ResNet}: \\ $224 \times 224$ pixels.}} &$256 \times 256$ pixels, 1:1 & $53.45 \pm 5.03 $ \\ \cline{2-3}
 &$128 \times 128$ pixels, 1:2 & $50.64 \pm 0.78 $ \\ \cline{2-3}
							 &$64 \times 64$ pixels, 1:4 & $57.03 \pm 15.86 $ \\ \cline{2-3}
 							 &$32 \times 32$ pixels, 1:8 & $52.48 \pm 1.30$ \\ \cline{2-3}
							&$16 \times 16$ pixels, 1:16 & $53.89 \pm 4.68$\\ \hline \hline

RBF kernel SVM ($C=100$, $\gamma=0.1$) 
using nuclei features (described in Table~\ref{tab:PCa_features}).
&$256 \times 256$ pixels, 1:1 & $44.39 \pm 21.55 $ \\ \hline \hline

Combination of nuclei features with DL features using MLP
 (Not including ResNet-22)
& All scales from $256 \times 256$ pixels to 
$16 \times 16$ pixels. 
& \textbf{85.93} $\pm$ \textbf{7.54} \\ \hline
\end{tabular}
\end{table}

\section{Conclusion}
Pre-trained `VGG16', `VGG19', `ResNet-50', `DenseNet-121', `DenseNet-169', and `Inception-v3' were fine-tuned and tested to assess the 
possibility of using transfer learning for cribriform pattern detection. The performances of these models in their individual and 
combined capacity were assessed. Various hand-crafted nuclei features were also designed and tested for cribriform pattern detection. 
Some of these nuclei feature has been successful in prostate cancer grading which is easier problem when compared to cribriform pattern 
detection. Cribriform patterns are one of patterns in high grade prostate cancer regions. Our ÒNon-cribriformÓ labelled images include 
various high grade PCa regions which appear similar to cribriform pattern w.r.t. nuclei texture and clustering. The fine-tuned DL 
models were able to correctly identify cribriform pattern as they were able to use the information not limited to just nuclei texture 
and location. The detection results at various scales using DL models were analysed and combined with nuclei features using MLP with 
improved performance. The cribriform detection results are promising and can be treated as a base-line for future projects. The current 
dataset includes images from Gleason pattern 3, Gleason pattern 4, and HG-PIN regions with color variations. Future studies should 
include cribriform pattern images from all possible sources and various color variations encompassing multiple patient information.

%\subsection*{Disclosures}
%
%The authors, Malay Singh, Emarene Mationg Kalaw, Wang Jie, 
%Mundher Al-Shabi, Chin Fong Wong, Danilo Medina Giron, Kian-Tai Chong,
% Maxine Tan, Zeng Zeng, and Hwee Kuan Lee have no financial disclosures and conflicts of interest.
 
%Conflicts of interest should be declared under a separate header. If the authors have no relevant financial interests in the manuscript and no other potential conflicts of interest to disclose, a statement to this effect should also be included in the manuscript.

\subsection*{Acknowledgments}
%This unnumbered section is used to identify those who have aided the authors in understanding or accomplishing the work presented and to acknowledge sources of funding. 

This work was supported in parts by the Biomedical Research Council of A*STAR (Agency for Science, Technology and Research), Singapore; 
Science and Engineering Research Council of A*STAR, Singapore; National University of Singapore, Singapore; Department of Pathology at 
Tan Tock Seng Hospital, Singapore;  Mount Elizabeth Novena Hospital,  Singapore; Farrer Park Hospital, Singapore;
University of Queensland, Australia; Monash University Malaysia, Malaysia; and Singapore-China NRF-Grant (No. NRF2016NRF-NSFC001-111).

%%%%% References %%%%%
\newpage
\bibliography{report} % bibliography data in report.bib
\bibliographystyle{ieeetr}

\end{document}